\documentclass[a4paper,10pt,onecolumn,notitlepage,final]{article}
\usepackage{graphicx}
\usepackage{bm}
\usepackage{booktabs}

\begin{document}
\title{An agent-based model for interrelation between COVID-19 outbreak and economic activities}
\maketitle
\author{Takeshi Kano$^{1}$$^{\ast}$, Kotaro Yasui$^{1,2}$, Taishi Mikami$^{1}$, Munehiro Asally$^{3,4,5}$, and Akio Ishiguro$^{1}$\vspace{5mm}\\ 
$^{1}${\em{Research Institute of Electrical Communication, Tohoku University, 2-1-1 Katahira, Aobaku, Sendai 980-8577, Japan}}\\
$^{2}${\em{Frontier Research Institute for Interdisciplinary Sciences, Tohoku University, Aramaki aza Aoba 6-3, Aoba-ku, Sendai 980-8578, Japan}}\\
$^{3}${\em{School of Life Sciences, University of Warwick, Coventry, CV4 7AL, UK}}\\
$^{4}${\em{Warwick Integrative Synthetic Biology Centre, University of Warwick, Coventry, CV4 7AL, UK}}\\
$^{5}${\em{Bio-Electrical Engineering Innovation Hub, University of Warwick, Coventry, CV4 7AL, UK}}\\
\thanks{$^\ast$Contact author. Tel: +81-22-217-5465. Email: tkano@riec.tohoku.ac.jp}

\begin{abstract}
As of July, 2020, acute respiratory syndrome caused by coronavirus COVID-19 is spreading over the world and causing severe economic damages. While minimizing human contact is effective in managing the outbreak, it causes severe economic losses. Strategies solving this dilemma by considering interrelation between the spread of the virus and economic activities are in urgent needs for mitigating the health and economic damage. Here we propose an abstract agent-based model for the outbreak of COVID-19 in which economic activities are taken into account. The computational simulation of the model recapitulated the trade-off between health and economic damage associated with lockdown measures. Based on the simulation results, we discuss how macroscopic dynamics of infection and economy emerge from the individuals’ behaviours. We believe our model can serve as a platform for discussing solutions to the abovementioned dilemma.
\end{abstract}


\section{Introduction}
The acute respiratory syndrome caused by a coronavirus, COVID-19, first reported in Wuhan in December 2019 [1--4], has spread over the world causing severe health and economic damages. To date (July 21th, 2020), more than 14 million people have been tested positive for the disease and more than 610 thousand people have died due to COVID-19 [5]. From the economic aspect, many people have suffered from economic losses and lost their jobs [6]. It should be noted here that there is a dilemma between mitigation of the spread of COVID19 and reduction of economic losses: namely, the strategies for decreasing contacts between humans such as draconic lockdown and social distancing hinder normal economic activities. Hence, it is an urgent issue to solve this dilemma and to find strategies to mitigate outbreak minimizing economic losses.

To date, many mathematical models for the spread of infectious diseases have been proposed using differential equations [7--10] and agent-based models [11--13]. Recently, mathematical models for the spread of COVID-19 have been proposed from various contexts, e.g. [14--23]. These studies solely modelled the spread of infectious diseases, yet they did not describe economic activities mathemtaically. Other studies focused on economic impacts of COVID-19 [24--28], some of which considered details of economic processes and predicted economic impact under realistic assumptions [24--26]; thus, they are complex and difficult to capture the essence. Meanwhile, the others estimated the economic impact using a simple model based on differential equations [27,28]. While these models describe the economic effects at a population level, how individual’s behaviour affects the macroscopic dynamics of infection and economy remains unclear. A definitive mathematical model that captures the essential mechanism of interrelation between the spread of the virus and economic activities at an individual level could fill this gap. 

In this study, we propose a simple mathematical model that considers both infections of COVID-19 and economic activities. Our aim is to extract the essence of the relation between the spread of COVID-19 and the economic activities, rather than making a quantitative and accurate prediction of infection and economy. Moreover, we are interested in how macroscopic dynamics of infection and economy emerge from individual’s behaviours, rather than macroscopic description through coarse graining. Hence, we propose a highly abstract and simple agent-based model without employing detailed and realistic assumptions. More specifically, we propose a cellular automaton model in which mobile agents with the internal states regarding infection and economy interact with others and update their internal states. We demonstrate through simulations how the health and economic state evolves depending on parameters. Based on the results, we discuss the effect of lockdown.

\section{Model}
\begin{figure*}[t]
    \begin{center}
         \includegraphics[width=0.9\hsize]{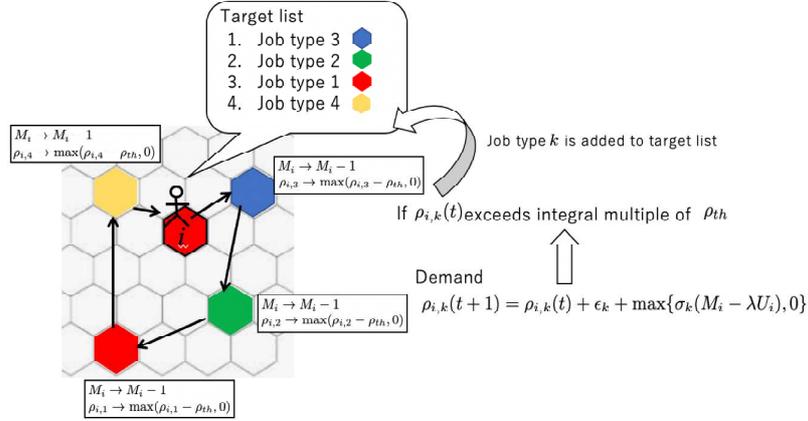}
    \end{center}
    \caption{Outline of the proposed model. When $\rho_{i,k}(t)$ exceeds integral multiple of $\rho_{th}$, job type $k$ is added to the bottom of the target list of agent $i$. Agent $i$ visits the home of the nearest agent among agents whose job type corresponds to the top of the target list. When it reached the home, it pays a unit amount money to buy the goods. Then, since the demand of agent $i$ is satisfied, $\rho_{i,k}$ decreases by $\rho_{th}$. The top item in the target list is removed, and the other items in the target list move up. Then, agent $i$ visits the next target. When there is no item in the target list, agent $i$ goes back and stay its home. 
     }
    \label{fig:model}
\end{figure*}

\begin{figure}[t]
    \begin{center}
         \includegraphics[width=0.75\hsize]{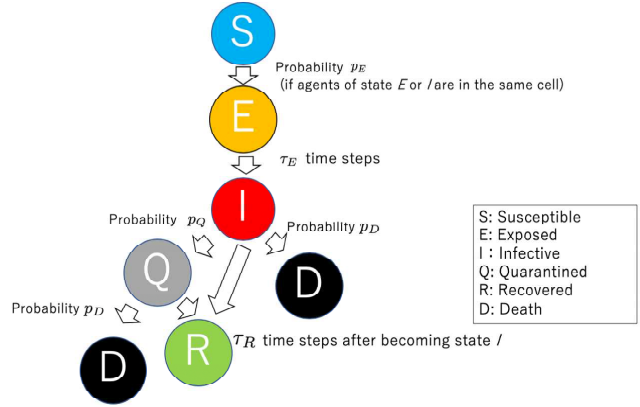}
    \end{center}
    \caption{Rule for the transition of the health state.
     }
    \label{fig:model-infection}
\end{figure}

\begin{figure}[t]
    \begin{center}
         \includegraphics[width=0.7\hsize]{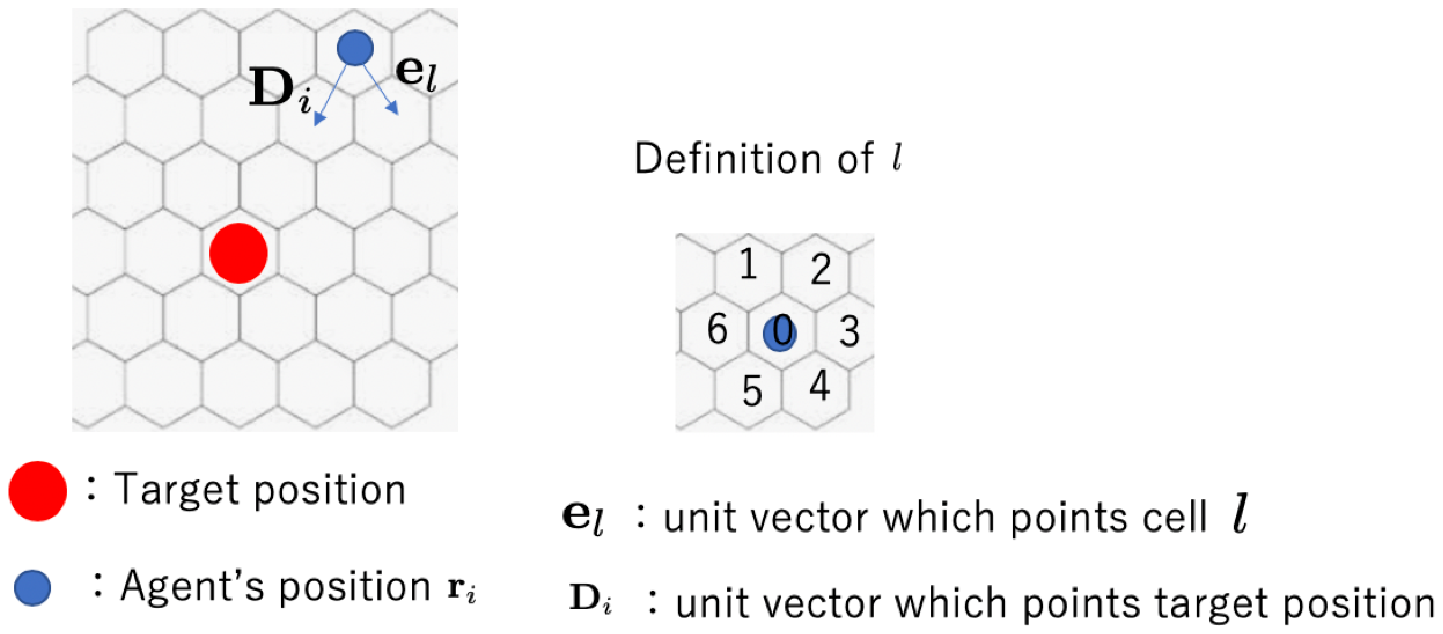}
    \end{center}
    \caption{Definition of $\mathbf{D}_i$ and $\mathbf{e}_l$.
     }
    \label{fig:model-movement}
\end{figure}

\subsection{Overview of the proposed model}
We consider a cellular automaton model in which hexagonal cells are aligned regularly on a two-dimensional plane (Fig. 1). $N$ agents are on the plane. Each agent has a home cell, and can stay at home or move to its adjacent cell every time step. Each agent has a health state $State_i$ and money $M_i$, which are updated through interaction with other agents. Agents die when they do not recover after infection or when their money becomes zero.

Agents have their own business. We assume that agent $i$ sells goods to agent $j$ when agent $j$ visits the home cell of agent $i$. Prime costs of goods are not considered. There are several types of businesses, which range from selling commodities to selling luxury goods. Each agent chooses one of them, and it does not change temporarily. Each agent has demands for goods. When the demand exceeds a certain threshold, the agent goes out to buy them; it pays money when it reached its home of an agent who sells the needed goods (Fig. \ref{fig:model}). When experiencing lockdown under outbreak, agents do not tend to demand luxury goods, while they demand commodities as usual. In the following subsections, we describe the details of the proposed model.

\subsection{Model of infection}
We model the spread of COVID-19 by drawing inspiration from a spatial susceptible-exposed-infectious-recovered (SEIR) model [13]. Each agent has a health state $State_i$, which can take a susceptible state (S), an exposed state (E), an infectious state (I), a quarantine state (Q), a recovered state (R), and a death state (D). 

The state changes according to the rule shown in Fig. \ref{fig:model-infection}. When an agent with state $S$ is in a cell occupied by an agent with state $E$ or $I$, $State_i$ changes from $S$ to $E$ with probability $p_E$ every time step. Here, although the previously proposed SEIR model [13] assumed that only agents with state $I$ have infectability, we assumed that agents with state $E$ in addition to state $I$ have infectability since asymptomatic patients of COVID-19 have infectability [29]. When $\tau_E$ time steps have passed after the transition from state $S$ to $E$, the state changes to state $I$. Agents with state $I$ changes to state $Q$ and $D$ with probability $p_D$ and $p_Q$, respectively, every time step. Agents with state $Q$, which represents hospitalized patients, do not have infectability and stop any economic activity described in the next subsection. Agents with state $Q$ also die with probability $p_D$ every time step. Regardless of whether quarantined or not, the state changes to $R$ when $\tau_R$ time steps have passed after the transition from state $E$ to $I$. 

\subsection{Model of economic activity}
Each agent has money $M_i$. The money $M_i$ decreases and increases when agent $i$ buys goods and when other agents buy agent $i$'s goods, respectively. In addition to buying and selling, high-income persons pay high taxes while low-income persons receive public assistance in real societies. Thus, we describe the time evolution of $M_i$ as
\begin{equation}
M_i(t+1) =M_i(t)+q_{in,i}(t)-q_{out,i}(t)+K(\hat{M}-M_i(t)),\label{M}
\end{equation}
where $K$ and $\hat{M}$ are positive constants, and $q_{in,i}(t)$ and $q_{out,i}$(t) denote the amount of money increased by selling and that decreased by buying, respectively. The fourth term on the right-hand side represents the effect of income redistribution: high-income persons pay high taxes and low-income persons receive public assistance. Agent $i$ dies when $M_i(t)$ becomes zero.

Each agent chooses one of $m$ job types, which does not change temporarily. Agent $i$'s demand for goods produced by job type $k$ ($i=1,2,\cdot\cdot\cdot, N$ and $k=1,2,\cdot\cdot\cdot, m$) is denoted by $\rho_{i,k}$. The time evolution of $\rho_{i,k}$ is described as
\begin{equation}
\rho_{i,k}(t+1)=\rho_{i,k}(t)+\epsilon_k+\max\{\sigma_k (M_i-\lambda U_i),0\},\label{rho}
\end{equation}
where $\epsilon_k$, $\sigma_k$, and $\lambda$ are positive constants, and $U_i$ represents the degree of an outbreak, which formulation is described below. Parameter $\epsilon_k$ represents the increase rate of the demand which does not depend on the money agent $i$ owns or the degree of the outbreak. In contrast, $\sigma_k$ represents the increase rate of the demand which is affected by the money agent $i$ owns or the degree of the outbreak. $\epsilon_k$ and $\sigma_k$ are set to be large and small, respectively, when job type $k$ treats commodities, while vice versa when job type $k$ treats luxury goods. The third term on the right-hand side of Eq. (\ref{rho}) means that the demand increases rapidly when agent $i$ has much money, yet the increase stops under the outbreak. Parameter $\lambda$ represents the degree of a lockdown under the outbreak.

The degree of outbreak $U_i$ is formulated as 
\begin{equation}
U_i(t+1)=(1-\kappa)U_i(t)+\kappa n_i(t),\label{U}
\end{equation}
where $n_i$ is the number of agents with state $Q$ within the radius $r$ from agent $i$, which represents the number of patients monitored in the residential area. Here, only the number of quarantined agents is counted, based on the assumption that people cannot notice that people who are exposed to the virus but not hospitalized (state $E$ and $I$) are actually exposed to the virus. Eq. (\ref{U}) means that $U_i(t)$ increases/decreases followed by the increase/decrease of $n_i(t)$. $U_i(t)$ is updated rapidly when $\kappa$ is large. Thus, $\kappa$ characterizes how fast lockdown is performed in response to the spread of the virus.

Each agent has a target list, which represents the priority of purchasing goods (Fig. \ref{fig:model}). When the demand $\rho_{i,k}$ exceeds the integral multiple of $\rho_{th}$, job type $k$ is added to the bottom of the target list of agent $i$. Agent $i$ visits the home of the nearest agent among agents whose job type corresponds to the top of the target list. When it reached its home, it pays a unit amount of money to buy the goods. Here we simply assumed that this deal holds even when the selling agent is not at its home. Then, since the demand of agent $i$ is satisfied, $\rho_{i,k}$ decreases by $\rho_{th}$. The top item in the target list is removed, and the other items in the target list move up. Then, agent $i$ visits the next target. When there is no item in the target list, agent $i$ goes back and stay its home. 

The rule for the movement of each agent is described as follows (Fig. \ref{fig:model-movement}). First, the target position of agent $i$ is defined as the home of the nearest agent among agents whose job type corresponds to the top of the target list. When there is no item in the target list, the target position is defined as agent $i$'s home. A unit vector which points the target position from the current position of agent $i$ is denoted by $\mathbf{D}_i$. When the target position is identical to the current position, $\mathbf{D}_i=\mathbf{0}$. Agent $i$ moves to its adjacent cells or stays at the current position with the following probability $p_i(l)$:
\begin{eqnarray}
p_i(l)&=&\frac{1}{6}(1+\mathbf{e}_{l}\cdot\mathbf{D}_i),~~~~~(l=1,2,3,4,5,6)\nonumber\\
p_i(0)&=&0,\label{idou}
\end{eqnarray}
in the case of $\mathbf{D}_i\neq\mathbf{0}$, and 
\begin{eqnarray}
p_i(l)&=&0,~~~~~(l=1,2,3,4,5,6)\nonumber\\
p_i(0)&=&1,\label{stop}
\end{eqnarray}
in the case of $\mathbf{D}_i=\mathbf{0}$, where the definition of the subscript $l$ in $\mathbf{e}_l$ is as shown in Fig. \ref{fig:model-movement}. Thus, owing to the bias term $\mathbf{e}_{l}\cdot\mathbf{D}_i$ in Eq. (\ref{idou}), agent $i$ tends to approach the target position.

\section{Simulation Results}
We performed simulations of the proposed model. The number of cells along the horizontal and vertical direction was 43 and 50, respectively, and periodic boundary condition was adopted. Except for the experiment shown in Fig. 9, the total number of agents $N$ was set to 1000. Initial amount of money was set to $\tilde{M}$ for all agents. The homes of the agents were randomly placed without any overlap. Each agent was initially located at its home. Each trial was performed for 30000 time steps.

\begin{table}[b]
\caption{Parameter values employed in the simulations.}
\label{tab:parameter}
\small
\begin{center}
\begin{tabular}{|c|c|c|c|}
\hline
parameter & value & parameter &value \\
\hline
\hline
$N$ & 1000 &$R$ &  10  \\
\hline
$p_E$ & 0.012 & $p_D$ & 0.0001 \\
\hline
$p_Q$ & 0.005 & $\kappa$ & 0.0004\\
\hline
$\tau_E$ & 600 & $\tau_R$ & 1200\\
\hline
$\rho_{th}$ & 100.0  & $K$ & 0.0007\\
\hline
$\tilde{M}$ & 60& $\hat{M}$ & 60 \\
\hline
\end{tabular}
\end{center}
\end{table}

There were four job types in the simulation. The number of people for each job type was set to be identical. Parameters $\epsilon_k$ and $\sigma_k$ ($k=1,2,3,4$) were determined as 
\begin{eqnarray}
\epsilon_k&=&0.1(k-1)\nonumber\\
\sigma_k&=&\frac{0.5-\epsilon_k}{\tilde{M}}.\label{sigma}
\end{eqnarray}
Thus, persons whose job type number was large sold commodities while those whose job type number was small sold luxury goods. From Eqs. (\ref{rho}) and (\ref{sigma}), the increase rate of the demand $\rho_{i,k}$ does not depend on the job type when $M_i=\tilde{M}$. 

Non-dimensional parameters were used in the simulation. Since the proposed model is highly abstract, it is difficult to determine the parameters based on realistic data. However, the duration of latent period $\tau_E$ and that of infection $\tau_R$ were chosen to roughly mimic the property of COVID-19. Specifically, $\tau_E$ and $\tau_R$ were set to be 600 and 1200; these non-dimensional values correspond to 7 and 14 days, respectively, if each time step is rescaled to 16.8 minutes, and these values roughly agree with the property of COVID-19. The death probability $p_D$ was set so that the death rate becomes about 10\% if there was no lockdown, {\em i.e.}, $\lambda=0$. Economic parameters such as $\tilde{M}, \hat{M}, \rho_{th}$, and $K$ were determined so that death caused by economic loss becomes almost zero in the absence of outbreak but some people die due to economic loss under lockdown during an outbreak. The parameter values thus determined are shown in Table 1. These values were used when unspecified hereafter.  

\subsection{Results when no patient exists}
To capture the basic property of the proposed model, simulations were performed under the condition where no patient exists, {\em i.e.}, $State_i=S$ for all $i$. By removing the factor of infection from the model, we can easily understand the basic property of economic activity. In this experiment, the parameter $K$ was changed to investigate the effect of income redistribution by taxes and public assistances. The results were evaluated by the number of deaths caused by economic loss $D_{eco}$ and the variance of the amount of money $V[M]\equiv N^{-1}\sum_{i=1}^N(M_i-\bar{M})^2$, where $\bar{M}$ is the average of $M_i$. 

Supplementary videos 1--3 provided in [30] show the results when $K=0, 0.00007$, and 0,00014, respectively. Figure 4 shows the time evolution of $V[M]$ and $D_{eco}$ for several values of $K$.  It is found that $V[M]$ increase as the time passes. The increase rate is smaller as $K$ is larger (Fig. 4(a)). The number of deaths $D_{eco}$ increased as the time passes for small $K$, while it remains zero for large $K$ (Fig. 4(b)). This result is reasonable because frequent buying and selling events cause the distribution of $M_i$ to spread in a diffusive manner. As $K$ increased, the difference between the rich and the poor decreased, and thus, the number of deaths caused by economic loss decreased. 

\begin{figure}[t]
    \begin{center}
         \includegraphics[width=0.6\hsize]{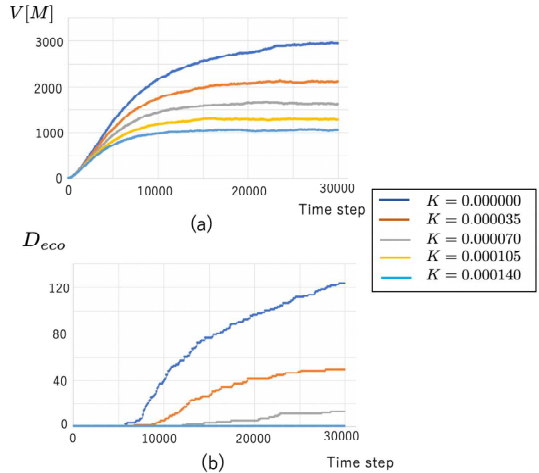}
    \end{center}
    \caption{Results when no patient exists.Time evolutions of (a) $V[M]$ and (b) $D_{eco}$ are shown for several values of $K$. In (b), lines for $K=0.000105$ and 0.000140 are overlapped.  
     }
    \label{fig:preliminary}
\end{figure}

\subsection{Effects of lockdown}
We performed simulations under the condition that the health state changed from $S$ to $E$ with the probability of 0.005 when the time step is 5000. Parameter $K$ was set to be 0.00007. Cases of $\lambda=0$, $120$, and $210$ were examined to investigate the effect of lockdown. 

Supplementary videos 4--6 provided in [30] show the results when $\lambda=0$, $120$, and $210$, respectively. Figs. 5 (a)--(c) show the time evolution of the number of agents with each health state. When $\lambda=0$, the infection spread rapidly and more than 80 \% of the agents were infected. The maximum number of the quarantined patients (state $Q$) was 278, and the number of deaths caused by infection $D_{inf}(t)$ finally reached 101 (Fig. 5(a)). As $\lambda$ increased, the number of patients and the deaths caused by infection decreased considerably (Figs. 5(b) and (c)). When $\lambda=210$, less than 20\% of the agents were infected (Fig. 5(c)).  

Next, we observed the economic tendency under these cases. The results were evaluated by the total number of deaths caused by economic loss and the total amount of money for each job type, denoted by $D_{eco, k}(t)$ and $T_k(t)$, respectively ($k=1,2,3,4$). Namely, $T_k(t)$ is given by
\begin{equation}
T_k(t)=\sum_{i\in {\rm job~type}~k}M_i(t).\label{Tk}
\end{equation}
The time evolutions of $D_{eco,k}(t)$ and $T_k(t)$ for the cases of $\lambda=0$, $120$, and $210$ are shown in Figs. 5(d)--(f) and (g)--(h), respectively.  For $\lambda=0$, only a few agents died because of economic loss (Fig. 5(d)). The total amount of money $T_k(t)$ was kept almost constant (Fig. 5(g)). When agents undergo lockdown, {\em i.e.}, $\lambda>0$, the economic gap among job types is generated. Specifically, $T_{k}(t)$ became small for small $k$ and vice versa for large $k$ followed by the outbreak. The economic gap is the highest around 15000 time steps, and it is mitigated gradually until around 30000 time steps (Figs. 5(h) and (i)). The number of deaths caused by economic loss is larger for small $k$ (Fig. 5(e)), and it is larger as $\lambda$ is larger (Fig. 5(f)). This result suggests that people who sell luxury goods suffer from economic loss caused by lockdown, while those who sell commodities do not. However, the persons who are suffering from economic loss can come back to their normal life after the outbreak is over. 

In summary, lockdown is effective for mitigating an outbreak, while it induces economic gap and increases the number of deaths caused by economic loss. 

\begin{figure*}[t]
    \begin{center}
         \includegraphics[width=1.0\hsize]{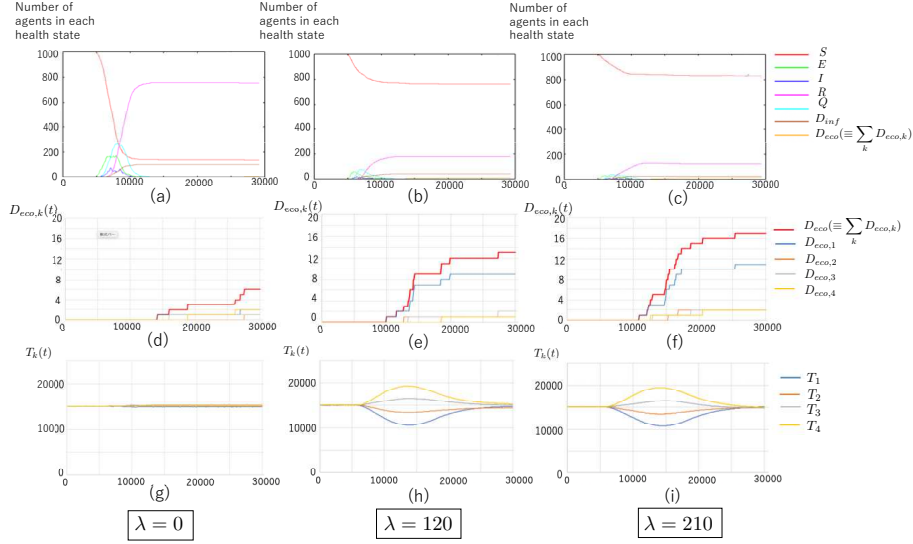}
    \end{center}
    \caption{Simulation results when patients appear at the time step 5000: (a)--(c) Time evolutions of the number of agents in each health state ($S$, $E$, $I$, $R$, $Q$, $D_{inf}$, and $D_{eco}$). (d)--(g) Time evolutions of $D_{eco,k}(t)$ and $D_{eco}(\equiv \sum_kD_{eco,k}(t))$. (h)--(i) Time evolutions of $T_k(t)$. $\lambda=0$ for (a)(d)(g), 120 for (b)(e)(h), and 210 for (c)(f)(i).  
     }
    \label{fig:lambda-dependence}
\end{figure*}

\subsection{Effect of other parameters}
\begin{figure}[t]
    \begin{center}
         \includegraphics[width=0.8\hsize]{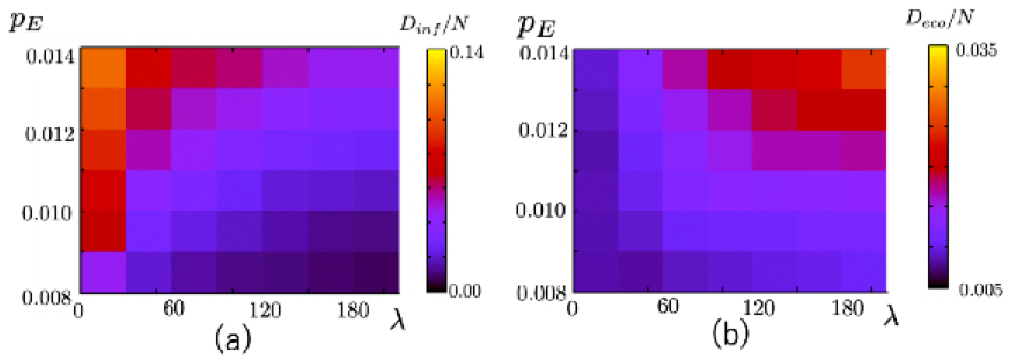}
    \end{center}
    \caption{Simulation results when $p_E$ and $\lambda$ are changed. $D_{inf}/N$ and $D_{eco}/N$ are shown in (a) and (b), respectively.
     }
    \label{fig:pE}
\end{figure}

\begin{figure}[t]
    \begin{center}
         \includegraphics[width=0.8\hsize]{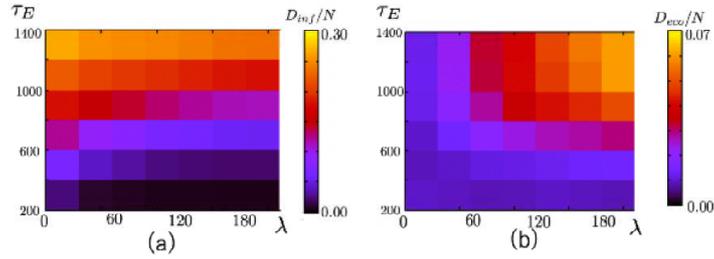}
    \end{center}
    \caption{Simulation results when $\tau_E$, $\tau_R$, and $\lambda$ are changed with satisfying $\tau_R=2\tau_E$. $D_{inf}/N$ and $D_{eco}/N$ are shown in (a) and (b), respectively. 
     }
    \label{fig:tauE}
\end{figure}

\begin{figure}[t]
    \begin{center}
         \includegraphics[width=0.77\hsize]{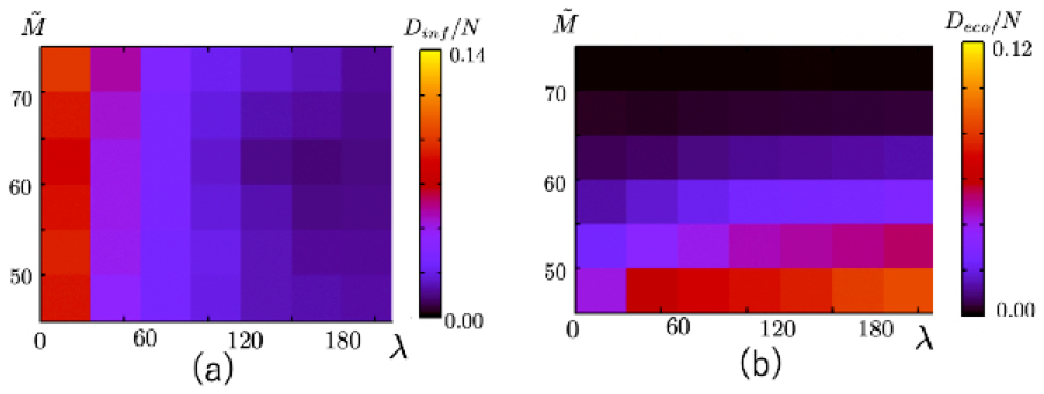}
    \end{center}
    \caption{Simulation results when $\tilde{M}$ and $\lambda$ are changed. $D_{inf}/N$ and $D_{eco}/N$ are shown in (a) and (b), respectively. 
     }
    \label{fig:tildeM}
\end{figure}

\begin{figure}[t]
    \begin{center}
        \includegraphics[width=0.78\hsize]{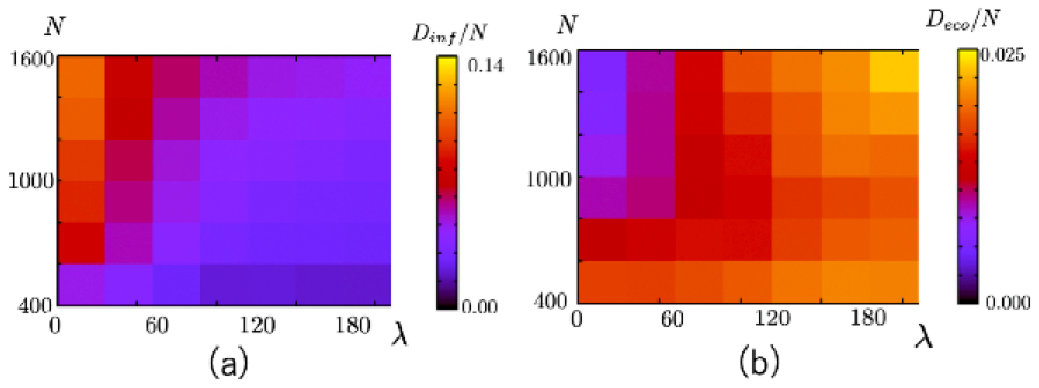}
    \end{center}
    \caption{Simulation results when $N$ and $\lambda$ are changed. $D_{inf}/N$ and $D_{eco}/N$ are shown in (a) and (b), respectively.  
     }
    \label{fig:N}
\end{figure}

\begin{figure}[t]
    \begin{center}
         \includegraphics[width=0.8\hsize]{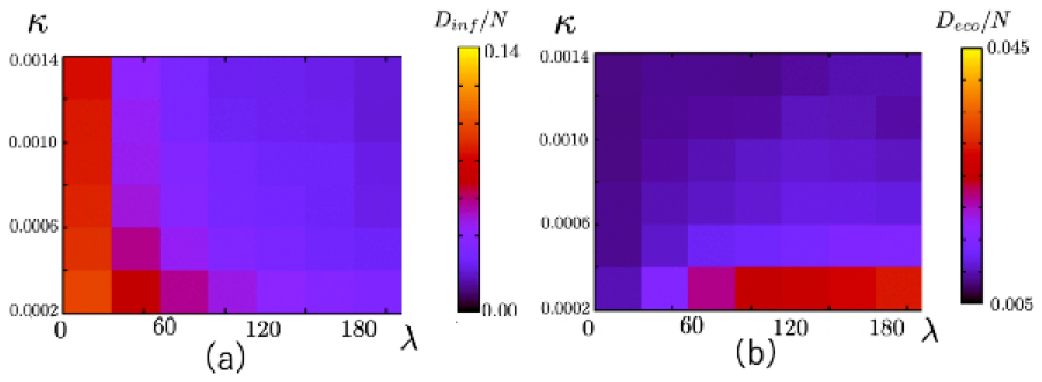}
    \end{center}
    \caption{Simulation results when $\kappa$ and $\lambda$ are changed. $D_{inf}/N$ and $D_{eco}/N$ are shown in (a) and (b), respectively.
     }
    \label{fig:kappa}
\end{figure}

For further understanding of the properties of the proposed model, we performed simulations by changing several parameters expected to affect resultant behaviours. Specifically, we changed the infectious rate $p_E$, the duration of infection $\tau_E$ and $\tau_R$, the amount of money $\tilde{M}$ and $\hat{M}$, the total number of agents $N$, and the quickness for the response to outbreak $\kappa$, since we expected that these parameters would affect infection and economic dynamics. For each parameter, we also changed $\lambda$ and made color maps. Other simulation conditions were the same as the previous subsection. Results were evaluated by the ratio of the number of deaths caused by infection $D_{inf}$ to the number of agents $N$ and that of the number of deaths caused by economic loss $D_{eco}$ to the number of agents $N$ at 30000 time step. Ten trials were performed for each parameter, and the average values of $D_{inf}/N$ and $D_{eco}/N$ are shown hereafter. In the following subsections, we will show the results when the parameters are varied.

\subsubsection{$p_E$ dependence}
Figure 6 shows the result when the infectious rate $p_E$ was changed. As is expected, the number of deaths is smaller as $p_E$ is smaller. When $p_E$ is large, $D_{inf}/N$ is large for small $\lambda$ while $D_{eco}/N$ is large for large $\lambda$. Namely, whereas infection spreads in the absence of lockdown, economic loss becomes large if lockdown is excessive. This trade-off becomes severer as the infectious rate is larger.

\subsubsection{$\tau_E$ and $\tau_R$ dependence}
Figure 7 shows the result when the duration of infection $\tau_E$ and $\tau_R$ were changed with holding the relation $\tau_R=2\tau_E$. When $\tau_E$ and $\tau_R$ are large, $D_{inf}/N$ is large for small $\lambda$ while $D_{eco}/N$ is large for large $\lambda$. Thus, the above-mentioned trade off becomes severer as the duration of infection is longer.

\subsubsection{$\tilde{M}$ and $\hat{M}$ dependence}
Figure 8 shows the result when the initial amount of money $\tilde{M}$ is changed. $\hat{M}$ in Eq. (\ref{M}) was also changed so that $\hat{M}=\tilde{M}$. The number of death by infection $D_{inf}/N$ is not much affected by $\tilde{M}$. However, $D_{eco}/N$ is larger as $\tilde{M}$ is smaller. This means that  people cannot tolerate lockdown if they do not have enough money.

\subsubsection{$N$ dependence}
Figure 9 shows the result when the total number of agents $N$ is changed. For small $\lambda$, $D_{inf}/N$ is larger as $N$ is larger, which indicates that infection tends to spread when population density is large. Meanwhile, $D_{eco}/N$ tends to increase as $\lambda$ increases.

\subsubsection{$\kappa$ dependence}
Figure 10 shows the result when the quickness for the response to outbreak $\kappa$ is changed. When $\kappa$ is small, the outbreak cannot be mitigated owing to the slow response. As a consequence, $D_{inf}/N$ tends to become large. Furthermore, since agents have to undergo lockdown for a long time, $D_{eco}/N$ tends to be also large. Thus, responding quickly is important for both mitigating outbreak and maintaining economic activities.

\section{Discussion and conclusion}
We proposed a simple and abstract mathematical model of COVID-19 outbreak with economic activities. Simulation results showed that lockdown measures enable mitigating outbreak, while it generates economic gap among job types. The reason why the economic gap generates can be explained as follows. The income of agents who sell luxury goods decreases during an outbreak because demands for luxury goods decrease owing to lockdown. Since they have to buy commodities as usual, they become poorer. In contrast, agents who sell commodities can get income as usual, yet they do not buy luxury goods owing to lockdown. As a consequence, they become richer. This suggests that in real societies, governments should make efforts to reduce the economic gap; otherwise, lockdown measures cannot be continued, which makes the mitigation of an outbreak difficult.

The result shown in Fig. 8 suggests that people cannot tolerate lockdown if their overall economic level is low. Thus, it is particularly harder for the communities with low economic power to mitigate an outbreak while maintaining economic activities. A possible solution to this is to respond to the spread of the virus as early as possible. As suggested from Fig. 10, if the response is done earlier (which corresponds to large $\kappa$ in our model), it is possible to stop an outbreak without causing severe economic damages. If a government has failed to respond quickly and the virus has spread, people should anticipate other possibilities such as the development of antiviral drugs and weakening of virus, which correspond to decreasing $p_E$ in our model.

Because we did not consider production activities and simply assumed that prime costs of goods are zero, our model is not suitable for making precise and quantitative predictions, especially on the economic impact. However, our model captures the essence of the interrelation between the spread of virus and economic activities, and thus, we believe that our model can become a platform for discussing strategies for mitigating outbreak while maintaining economic activities. Indeed, our model potentially has many possibilities in the future. For example, it may be possible to take the effect of long-distance movement with transportations like airplanes and trains into account. Defining the capacity of hospitals may enable us to discuss how to avoid overwhelming hospitals. It may be also interesting to introduce individual characters of each agent and to discuss how selfish persons affect spread of the virus and economic activities.

Moreover, beyond the outbreak of COVID-19, our model may share a common mechanism with other systems that can survive under harsh circumstances with limited resources, such as bacterial biofilms [31] and community of vampire bats [32]. Extracting the common mechanism may lead to establish a general design principle for artificial systems with high survivability.

\end{document}